\documentclass[preprint,
a4paper,
 superscriptaddress,
 nofootinbib,
 bibnotes,
 amsmath,amssymb, aps,
 prd,
floatfix,
]{revtex4-1}

\usepackage{url}
\usepackage{soul}
\usepackage{float}
\usepackage[utf8]{inputenc}
\usepackage{graphicx}
\usepackage[caption=false]{subfig}
\usepackage{dcolumn}
\usepackage{multirow}
\usepackage{tabularx}
\usepackage{bm}
\usepackage{hyperref}
\usepackage{amsmath}
\usepackage{amssymb}
\usepackage{comment}
\usepackage{color}
\usepackage{enumerate}
\usepackage{booktabs}
\usepackage{siunitx}
\usepackage{physics}

\newcommand{\tF}{\widetilde{F}}
\newcommand{\be}{\begin{eqnarray}}
\newcommand{\ee}{\end{eqnarray}}

\usepackage[usenames,dvipsnames,svgnames]{xcolor}

\begin{document}

\title{The Dark Matter Phonon Coupling}%

 \author{Peter Cox}%
\email{peter.cox@ipmu.jp}
\affiliation{Kavli Institute for the Physics and Mathematics of the Universe (WPI),  UTIAS, The University of Tokyo, Kashiwa, Chiba 277-8583, Japan}

 \author{Tom Melia}%
\email{tom.melia@ipmu.jp}
\affiliation{Kavli Institute for the Physics and Mathematics of the Universe (WPI),  UTIAS, The University of Tokyo, Kashiwa, Chiba 277-8583, Japan}

 \author{Surjeet Rajendran}%
\email{surjeet@berkeley.edu}
\affiliation{Department of Physics, University of California, Berkeley, California 94720, USA}
\affiliation{Department of Physics \& Astronomy, The Johns Hopkins University, Baltimore, MD  21218, USA}

\preprint{IPMU19-0065}%

\begin{abstract}{
Generically, the effective coupling between the dark matter and an atom scales with the number of constituents in the atom, resulting in the effective coupling being proportional to the mass of the atom.  In this limit, when the momentum transfer is also small, we show that the leading term in the scattering of a particle off the optical phonons of an array of atoms, whether in a crystal or in a molecule,  vanishes. Next-generation dark matter direct detection experiments with sub-eV energy thresholds will operate in a regime where this effect is important, and the suppression can be up to order $10^6$ over naive expectations. For dark matter that couples differently to protons and neutrons, the suppression is typically of order $10-100$ but can be avoided through a judicious choice of material, utilising  variations in nuclear ratios $Z/A$ to break the proportionality of the coupling to mass. We provide explicit illustrations of this effect by calculating structure factors for di-molecules and for the crystals NaI and sapphire. 

}
\end{abstract}

\maketitle

\section{Introduction}
\label{sec:intro}

The experimental endeavour to directly detect dark matter must confront a possible dark matter mass range spanning over fifty orders of magnitude (see {\it e.g.} ref.~\cite{Lin:2019uvt} for a recent review). Within the lowest mass regions, $10^{-22}\lesssim m_{\text{DM}}/\text{eV} \lesssim 10^2$, the dark matter oscillates as a coherent classical field, which can be leveraged in experiments that search for resonant effects (see {\it e.g.}~\cite{Graham:2015ouw}). For higher masses, such  effects are absent, and conventional WIMP searches rely on detection of energy deposited in a scattering event. Detector energy thresholds are being pushed lower, with current technology demonstrating sensitivity to around a few eV of energy deposit~\cite{Agnese:2018col,Crisler:2018gci}, probing $m_{\text{DM}}\gtrsim$\,MeV. New technologies are needed, and are being developed (see {\it e.g.}~\cite{Battaglieri:2017aum} for an overview), to probe the currently inaccessible mass region between $10^2\lesssim m_{\text{DM}}/\text{eV}\lesssim 10^6$. These scattering events have momentum transfer $q$ in the region of $\sim$\,0.1\,eV--1\,keV. A number of proposals exploit a dark matter interaction with phonons, as these are the relevant quanta at these low energy/momentum transfers~\cite{Hochberg:2015pha,Schutz:2016tid,Hochberg:2016sqx,Bunting:2017net,Hochberg:2017wce,PhysRevX.8.041001,Knapen:2017ekk,Griffin:2018bjn}; for crystal-based proposals~\cite{Bunting:2017net,Knapen:2017ekk,Griffin:2018bjn}, optical phonons are important as these have the correct kinematics to efficiently couple to light dark matter.

In this paper, we highlight a particular feature of dark matter--phonon interactions in target crystals or molecules in the $q\lesssim\,$keV window of low momentum transfer that, to the best of our knowledge, has not been pointed out in the literature. We consider two types of interaction. The first is scattering of a particle by an array of $N$ atoms via a potential of the form
\be \label{eq:potential}
  \mathcal{V} = \sum_{i=1}^{N} g_i V({\bf r} - {\bf r}_i ) \,,
\label{eq:scatteringpot}
\ee 
where ${\bf r}$ is the position of the incoming particle, ${\bf r}_i$ are the positions of the atoms,  $g_i$ is the coupling to the $i$th atom, and where $V({\bf r} - {\bf r}_i )$ can account for both long and short range interactions. The second is the interaction of a field ${\bf A}$ via the dipole operator
\be
D= -\sum_{i=1}^N g_i\,{\bf r}_i \cdot {\bf A} \,,
\label{eq:dipole}
\ee
where ${\bf A}$ can describe a vector field or the gradient of a scalar field and is treated as being constant in space and time, compared with the size of the system.
In both cases, we show that the leading order (proportional to $q^2)$ scattering off optical modes vanishes in the limit where the scattering particle couples to the target atoms proportional to their masses. That is, denoting the mass of the $i$th atom as $m_i$, the leading  term vanishes if $g_i = g \,m_i$ for all  $i$, and for some constant $g$. 

The result follows from conservation of momentum. 
 First, consider the transition matrix element describing scattering via the potential eq.~\eqref{eq:potential}, with momentum transfer ${\bf q}$ and where the crystal/molecule target goes from state $|\Phi_i\rangle$ to $|\Phi_f\rangle$:
\be
\langle \Phi_f| V({\bf q}) \sum_l g_l e^{i {\bf q}\cdot{\bf \hat{r}_l}} |\Phi_i \rangle \,,
\label{eq:matrixel}
\ee
with $V({\bf q})=\int d^3{\bf r'}\, e^{i\, {\bf q}\cdot{\bf r'}} V({\bf r'})$. For inelastic scattering in the $q\to 0$ limit we take the linear term in the expansion of the exponential. Setting $g_l = g \,m_l$, this becomes
\be
 i V({\bf q})\, g \,{\bf q}\cdot \,\langle \Phi_f|\left(\sum_l m_l {\bf \hat{r_l}}\right) |\Phi_i \rangle  = iV({\bf q})\, g \,{\bf q}\cdot \,\langle \Phi_f| \,{\bf \hat{R}} \,|\Phi_i \rangle  \,,
\label{eq:rzero}
\ee
where ${\bf \hat{R}}$ is the centre of mass (COM) coordinate of the target. Momentum conservation guarantees that the COM coordinate operator can never induce a transition between different internal states, and so the matrix element in eq.~\eqref{eq:rzero} is zero. 
The same argument clearly holds for transition elements involving the dipole operator in eq,~\eqref{eq:dipole}, $\langle \Phi_f|D|\Phi_i \rangle$. Note that we have assumed nothing about the internal states of the system, so the effect is general; we will however use the harmonic approximation in the next section, and explicitly show how these arguments work in that case.  

One should still ask: in what regime is the above low $q$ expansion valid?
Clearly, it applies whenever $1/q$ is larger than the size of the entire system, for example when scattering off a molecule. On the other hand, for scattering off the optical modes of a periodic lattice the relevant scale is in fact the size of the unit cell ($q\lesssim\,$keV). 
This can be understood as follows: first, for a periodic system we need only consider the matrix element in eq.~\eqref{eq:matrixel} with the sum restricted to be over a single unit cell; the matrix elements for atoms in other unit cells are related by a phase factor due to Bloch's theorem. We then apply the same argument as above: $\bf\hat{R}$ now becomes the COM coordinate of the unit cell and as such can never induce transitions involving optical phonons. Note that it can still induce transitions between acoustic phonons as these are translations of the unit cell; for these modes the relevant scale remains the total size of the system. However, the kinematic mismatch between the virial velocity of dark matter and the speed of sound in materials makes it difficult to efficiently excite acoustic phonons for light dark matter detection. 

This ``coupling-to-mass'' limit is a generic feature of dark matter interactions with atoms and molecules that are being searched for in proposed sub-eV crystal or molecule-based direct detection experiments (see refs.~\cite{Bunting:2017net,Knapen:2017ekk,Griffin:2018bjn, Green:2017ybv}). In these experiments, the dark matter is assumed to have some interaction with individual nucleons and electrons, for example, via couplings to their electric, baryon or weak charges.  At the low momenta ($\lessapprox$ keV) transferred in these collisions, the dark matter effectively couples coherently to the entire atom, resulting in an effective coupling that is typically {\it proportional} to the mass of the atom. 

We proceed more quantitatively by writing the coupling of the dark matter to the $i$th atom, postponing the treatment for coupling to electrons (which is at any rate strongly constrained~\cite{Green:2017ybv}) to section~\ref{sec:discussion}, as
\be
g_i &=&  g_p \, Z_i + g_n \, (A_i-Z_i)  \label{eq:gNtwo}\\
&=&  A_i \, \left( (g_p-g_n) \, \frac{Z_i}{A_i} + g_n\right) \,,
\label{eq:gN}
\ee
where $g_p$ is the coupling to protons, $g_n$ is the coupling to neutrons, and $Z_i$, $A_i$ are the proton number and atomic mass number, respectively. Considering first the case  $g_p=g_n=g$, we have coupling proportional to atomic mass number $g_i= g \, A_i$. Since the atomic mass number and the physical mass, $m_i$, of a nucleus differ due to binding energies and the proton/neutron mass difference, which are both MeV effects, we expect a deviation from the coupling-to-mass limit of order $\epsilon\sim$\,MeV/GeV\,$\sim10^{-3}$. We will see that the $q^2$ term in the scattering rate is proportional to $\epsilon^2$, such that for this case of coupling to baryon number, we expect higher-order $q^4$ terms to be dominant. While the formalism for the scattering of dark matter with phonons was discussed in detail in \cite{Griffin:2018bjn},  
this coupling-to-mass effect was missed in the analytic analysis; we emphasise that this is a cancellation that exists for arbitrary mass difference between atomic species, contrary to the claim in \cite{Griffin:2018bjn}.

Moving to the more general case where $g_p\ne g_n$, we see from eq.~\eqref{eq:gN} that deviations from coupling proportional to atomic mass number are characterised {\it solely} by the ratio $Z_i/A_i$. For instance, for molecules or crystals that consist of nuclei that all have an equal value of the ratio $Z_i/A_i$ (for many light elements $Z/A=1/2$), the coupling is again proportional to atomic mass number and, as above, we expect a huge suppression of the leading term in the scattering rate. For $g_p\ne g_n$, the rate can therefore be significantly enhanced by a choice of material that consists of nuclei with differing $Z_i/A_i$. 

We point out that there exist two important cases which are in practice far from the coupling-to-mass limit. The first is where the interaction of the dark matter with the Standard Model proceeds through a dark photon that is kinetically mixed with the photon. In this case, due to the nature of the Standard Model electric charges, the effective coupling of the dark photon to the crystal proceeds through an atomic/molecular electric dipole moment which is not proportional to the mass of the atom/molecule. The second case is spin dependent couplings, which  are also not typically proportional to the mass of the atom. In these cases, one could instead choose a material with equal $Z_i/A_i$ in order to suppress the neutrino scattering background.

We note that in the case of photons being absorbed or undergoing Raman scattering in crystals and molecules, it is known that suppressions of the leading term can also occur; certain phonon modes can be inactive, or `IR-silent' (see e.g.~\cite{FERRARO20031}). The mechanism by which this phenomena takes place, however, is based on lattice symmetry selection rules, and differs from the coupling-to-mass mechanism studied here.

The remainder of the paper proceeds as follows. In section~\ref{sec:couplingtomass}, we give a proof of the coupling-to-mass effect in general harmonic systems, for both scattering via eq.~\eqref{eq:scatteringpot} and absorption via eq.~\eqref{eq:dipole}; we also provide an interpretation using the Fr\"ohlich interaction. In section~\ref{sec:examples} we illustrate the effect via examples: scattering and absorption in diatomic molecules, and scattering in the crystals NaI and Al$_2$O$_3$ (sapphire). Section~\ref{sec:discussion} discusses the relevance of the effect for dark matter direct detection experiments. Details of the scattering formalism are included in appendix~\ref{sec:appa}.

\section{The coupling to mass effect in harmonic systems}
\label{sec:couplingtomass}

\subsection{Inelastic scattering}
Here we prove that the one-phonon inelastic structure factor that describes scattering via the potential eq.~\eqref{eq:scatteringpot} exhibits the coupling-to-mass effect for any harmonic system. We start with the Hamiltonian describing the interaction of the atoms in a crystal at harmonic order,
\be
  H = \frac{1}{2} \sum_{\alpha,\beta=1}^3 \sum_{i,j=1}^N \left( p^\alpha_i A^{\alpha\beta}_{ij}p^{\beta}_j + x^\alpha_i F^{\alpha\beta}_{ij} x^{\beta}_j \right) \,,
\label{eq:ham}
\ee
where the sum on $\alpha,\beta$ runs over spatial dimensions, and $i,j=1\ldots N$ are site indices which run over all $N$ atoms in the crystal; $x_i^\alpha$ denotes the displacement of the $i$th atom from its equilibrium position, $X_i^\alpha$. The mass matrix $A^{\alpha\beta}_{ij}=\delta^{\alpha\beta}\text{diag}(1/m_1,..,1/m_N)_{ij}$, where $m_i$ is the mass of the $i$th atom. The force constants matrix, $F^{\alpha\beta}_{ij}$, is the second-order expansion of the crystal potential $U(x_1,\ldots,x_N)$,
\be
  F^{\alpha\beta}_{ij} = \frac{\partial^2U}{\partial x^\alpha_i \partial x^\beta_j} \bigg|_{x=0} \,,
\ee
which is symmetric in $\alpha,\beta$ and $i,j$; further, momentum conservation $[H,\sum_i p_i^\alpha]=0$ implies the following property,
\be
  \sum_{j=1}^N F^{\alpha\beta}_{ij} =0 \,.
\label{eq:momconsa}
\ee

To diagonalise the system, we first rescale the Hamiltonian, eq~\eqref{eq:ham}, sending $ p_i^\alpha = \sqrt{m_i}\,\tilde{p}_i^\alpha$ and $x_i^\alpha = \tilde{x}_i^\alpha /\sqrt{m_i}$. This gives
\be
  H &=&  \frac{1}{2} \sum_{\alpha,\beta} \sum_{i,j} \left( \tilde{p}^\alpha_i \tilde{p}^\beta_j \delta^{\alpha\beta}\delta_{ij} + \tilde{x}^{\alpha}_i\, \widetilde{F}^{\alpha\beta}_{ij} \,\tilde{x}^\beta_j \right) 
\,,
\ee
where
\be
  \widetilde{F}^{\alpha \beta}_{ij}=  F^{\alpha\beta}_{ij} \frac{1}{\sqrt{m_i m_j}} \,.
\ee
Note that the property of $F$ that followed from momentum conservation, eq.~\eqref{eq:momconsa},  implies that
\be
  \sum_{j=1}^N \tF^{\alpha\beta}_{ij}\sqrt{m_j} =0 \,.
\label{eq:momconsb}
\ee

The next step is to finally diagonalise the system, {\it i.e.} find the eigenvectors and eigenvalues of $\widetilde{F}$. It is easiest to visualise by combining the $\alpha$ and $i$ indices into one single $3N$ dimensional index (and similarly for $\beta$, $j$) in the following way,
\be
  (\tF) = \left( \begin{array}{ccc} 
  (\tF^{xx}) & (\tF^{xy})& (\tF^{xz}) \\
  (\tF^{yx}) & (\tF^{yy})& (\tF^{yz}) \\
  (\tF^{zx}) & (\tF^{zy})& (\tF^{zz})
  \end{array} \right)\,.
\ee
We denote the $3N$ eigenvectors of this matrix by ${\bf v}_a$, with corresponding eigenvalues $\omega_a^2$. As a result of eq.~\eqref{eq:momconsb}, we see that the 3 dimensional subspace of eigenvectors with zero eigenvalue---the acoustic vibrational modes---is spanned by the eigenvectors
\be
  {\bf v}_1=\mathcal{N}\left(\begin{array}{c} {\bf A}^{-1/2} \\ \mathbf{0}_N \\ \mathbf{0}_N \end{array} \right) \,,~~{\bf v}_2=\mathcal{N}\left(\begin{array}{c} \mathbf{0}_N\\ {\bf A}^{-1/2} \\  \mathbf{0}_N \end{array} \right)\,,~~{\bf v}_3=\mathcal{N}\left(\begin{array}{c} \mathbf{0}_N \\\mathbf{0}_N \\ {\bf A}^{-1/2}  \end{array} \right) \,,
\ee
where ${\bf A}^{-1/2} = (\sqrt{m_1},\ldots,\sqrt{m_N})^T$, and $\mathcal{N}=(\sum_{i=1}^N m_i)^{-\frac{1}{2}}$. Switching back to $\alpha$ and $i$ notation these acoustic eigenvectors are 
\be
  (v_1)^\alpha_i = \mathcal{N}\delta^{\alpha x} (A^{-1/2})_{i} \,,\quad
  (v_2)^\alpha_i = \mathcal{N}\delta^{\alpha y} (A^{-1/2})_{i} \,,\quad
  (v_3)^\alpha_i = \mathcal{N}\delta^{\alpha z} (A^{-1/2})_{i} \,.
\ee

Now, consider an incoming particle that scatters off the atoms in the crystal via a potential of the form in eq.~\eqref{eq:potential}. The rate for inelastic scattering that (de-)excites the mode with eigenvector ${\bf v}_a$ is proportional to the form factor (see appendix)
\be \label{eq:formfactor}
  \mathcal{F}_a({\bf q}) = \sum_{\alpha=1}^3 \sum_{l=1}^N \frac{g_l}{\sqrt{m_l}} q^\alpha  (v_a)^{\alpha}_l e^{i\,{\bf q}\cdot{\bf X}_l} e^{-W_l({\bf q})} \,,
\ee
where $W_l({\bf q})$ is the Debye-Waller factor. The leading term in the small-$q$ expansion of this expression is
\be
  \mathcal{F}_a({\bf q}) = \sum_\alpha \sum_l  \frac{g_l}{\sqrt{m_l}} q^\alpha  (v_a)^{\alpha}_l +\ldots \,.
\label{eq:leading}
\ee
One can  see that if the couplings are proportional to the masses, $g_l = g m_l$ for some $g$, only scattering into the acoustic modes is non-zero, by orthogonality of the eigenvectors. Explicitly, we can write the term appearing in eq.~\eqref{eq:leading} as
\be
 \sum_\alpha \sum_l  \frac{g m_l}{\sqrt{m_l}} q^\alpha  (v_a)^{\alpha}_l &=& g \sum_\alpha \sum_l \Big(q^x (A^{-1/2})_l \delta^{x\alpha} +  q^y (A^{-1/2})_l \delta^{y\alpha}  +  q^z (A^{-1/2})_l  \delta^{z\alpha} \Big) (v_a)^\alpha_l  \nonumber \\
 &=& \frac{g}{\mathcal{N}} (  q^x  {\bf v}_1  +q^y  {\bf v}_2+q^z  {\bf v}_3 ) \cdot {\bf v}_a \,.
  \label{eq:orthog}
\ee
Thus, unless $a=1,2$ or $3$ this vanishes by orthogonality.

Parameterising a deviation from the coupling-to-mass case by $g_l= g m_l(1+\epsilon c_l)$ (where $c_l$ are order one numbers to encode different deviations for different atoms), we see that the cross-section for scattering into optical modes ($\sigma\,\propto|\mathcal{F}|^2$) is proportional to $\epsilon^2$ at small $q$.

\subsection{Absorption via a dipole interaction}

With the above formalism, we can now see the mechanics of the coupling-to-mass effect in dipole transitions, the rate of which is proportional to
\be
\left|\langle \Phi_f | \sum_l g_l \,({\bf X}_l+{\bf x}_l) \cdot {\bf A}  | \Phi_i \rangle \right|^2 \,.
\ee
Absorption into the mode with eigenvector ${\bf v}_a$, is proportional to the form factor, $|\mathcal{F}_a({\bf A})|^2$, where
\be
\mathcal{F}_a({\bf A}) =\sum_\alpha \sum_l \frac{g_l}{\sqrt{m_l}} A^\alpha ( v_a)^\alpha_l \,.
\label{eq:formfactorabsorb}
\ee
It is clear that we can perform the same manipulations as in eq.~\eqref{eq:orthog} to show this vanishes in the $g_l\propto m_l$ limit.

\subsection{The Fr\"ohlich interaction}

The Fr\"ohlich interaction provides another way of describing low momentum transfer scattering of a particle by phonons in a periodic crystal; it is particularly useful to succinctly capture electromagnetic screening effects in the case of scattering via a photon. Because it describes the same physics as the structure factor approach at low $q$, it should also exhibit the coupling-to-mass effect. The matrix element for electromagnetic scattering by the eigenmode ${\bf v}_a$ at low momentum transfer, ${\bf q}\to {\bf 0}$, is~\cite{doi:10.1080/00018735400101213,PhysRevB.13.694,PhysRevLett.115.176401},
\be
\mathcal{M}_{a,{\bf q}} \propto \sum_{\alpha,\beta=1}^3\sum_{l=1}^{N_b} \frac{e}{\sqrt{m_l}} \frac{ q^{\alpha}Z^{*\,\alpha \beta}_{l} (v_a)^\beta_l}{\left(\sum_{\gamma,\delta}q^\gamma \epsilon_\infty^{\gamma \delta} q^\delta\right)} \,,
\label{eq:frohlich}
\ee
where $N_b$ is the number of atoms in the unit cell,  $e$ is the electron electromagnetic charge, and $\epsilon_\infty^{\gamma\delta}$ is the dielectric permittivity tensor. The quantities $Z^{*\alpha\beta}_{i}$ are the Born effective charges. For electromagnetism, they satisfy the sum rule 
\be \label{eq:effcharge_sumrule}
  \sum_{l=1}^{N_b} Z^{*\alpha\beta}_{l} = 0 \,,
\ee
which guarantees charge neutrality within the unit cell. Note how this follows from requiring that this matrix element is zero for scattering into the non-dipole ({\it i.e.} acoustic) modes with eigenvectors $(v_a)_l^\beta$, $a=1,2$ or $3$; the $1/\sqrt{m_l}$ cancels the $\sqrt{m_l}$ in the $v_a$, such that the only $l$ dependence is contained in the $Z^{*\,\alpha\beta}_{l}$.

The matrix element eq.~\eqref{eq:frohlich} reduces to the form obtained in the structure function approach to low $q$ scattering via a light mediator (which provides a factor of $1/q^2$) upon setting $e Z^{*\,\alpha\beta}_{l} = \delta^{\alpha\beta} g_l$, $\epsilon_\infty^{\gamma \delta}=\delta^{\gamma\delta}$, and where $\sum_l g_l\ne0$ in general.

The coupling-to-mass effect is indeed apparent in the form of eq.~\eqref{eq:frohlich}: if $Z_l^{*\,\alpha\beta} = g^{\alpha \beta} m_l$ (where now $g^{\alpha \beta}$ is a constant `couplings  tensor')  it again follows by the  mechanism of eq.~\eqref{eq:orthog} that scattering into optical modes vanishes. 

However, for the case of electromagnetism, the additional physical requirement of charge neutrality and the resulting sum rule, eq.~\eqref{eq:effcharge_sumrule}, ensures deviation from the coupling-to-mass limit. That is, dark photons that kinetically mix with the Standard Model photon give rise to interactions (scatterings, or absorption, via an equivalently screened version of eq.~\eqref{eq:formfactorabsorb}) that are away from the coupling-to-mass limit.

\section{Examples}
\label{sec:examples}

\subsection{Scattering and absorption in a diatomic molecule}

The simplest system that exhibits the coupling-to-mass effect is a diatomic molecule, composed of two atoms of mass $m_1$ and $m_2$. We model it as a one dimensional harmonic system,  the Hamiltonian of which takes the form in eq.~\eqref{eq:ham}. We will discuss rotational modes and anharmonic corrections to the potential shortly.  

In this case it is trivial to diagonalise the system, and the form factor for scattering into the optical mode via a potential of the type in eq.~\eqref{eq:potential} is simply
\be
| \mathcal{F}(q) |^2 = \frac{q^2}{M} \left( \frac{g_1^2 m_2}{m_1} + \frac{g_2^2 m_1}{m_2} - 2 g_1 g_2 \cos q a \right ) \,,
\label{eq:formfacmolecule}
\ee
where $a=|X_2 - X_1|$ is the equilibrium inter-atomic distance, and $M=m_1+m_2$. 
We have neglected the Debye-Waller factor since for low momentum transfer $W_l\approx1$ with corrections suppressed by $q^2/(\mu \omega)$, where $\mu=m_1 m_2/M$ is the reduced mass of the system. Similarly, the form factor for absorption in the presence of the field $A$ via eq.~\eqref{eq:dipole} is
\be
| \mathcal{F}(A) |^2 = \frac{A^2}{M m_1 m_2} \left( g_1m_2 -g_2 m_1 \right )^2 \,.
\ee
The analysis in this case is a trivialisation of the one for scattering and we do not explicitly present it.

We parameterise the couplings as $g_1=g m_1(1+\epsilon/2)$ and $g_2=g m_2(1-\epsilon/2)$, where $\epsilon$ provides the deviation from the coupling-to-mass limit. Expanding the form factor eq.~\eqref{eq:formfacmolecule} in the limit of small $\epsilon$ and small $qa$, we find
\be
  | \mathcal{F}(q) |^2 = q^2g^2 \mu \left( \epsilon^2 + (qa)^2 + \ldots \,\right) \,,
\label{eq:molform}
\ee
 where we drop higher order terms in $(qa)$ or $\epsilon$.
As expected from the  general arguments laid out in the introduction, we see that the leading $q^2$ term is suppressed by a factor of $\epsilon^2$. 

The $\epsilon^2$ term dominates the rate for $q\lesssim \epsilon/a$. In the other regime where the $q^4$ term in eq.~\eqref{eq:molform} dominates, one should also calculate other  contributions to the rate coming from the two-mode transitions, which may be important since $q^2/\mu w$ can be of order $(qa)^2$.

We finally turn to a discussion of rotational modes and anharmonic corrections. An inclusion of these renders the effectively one-dimensional example above a more realistic description of dark matter scattering off a di-molecule. 
Rotational modes essentially factor from the vibrational ones described above (see {\it e.g.} ref.~\cite{PhysRevX.8.041001}), and represent a hyper-fine splitting to the above analysis, such that they can be safely neglected at this level of discussion of the scattering rate. Anharmonic corrections, on the other hand, are more important; they can be analysed to a good approximation using the Morse inter-atomic potential~\cite{PhysRev.34.57} and provide contributions at relative order $\sim 10^{-2}$ to scattering rates  (again, see {\it e.g.} ref.~\cite{PhysRevX.8.041001}). However, note that even in an analysis that includes the full potential, the leading term in $q^2$ is still suppressed by $\epsilon^2$ following the general arguments from the introduction.

\subsection{Scattering in crystals}

Next, we provide two examples which illustrate the coupling-to-mass effect in crystals. 
Specifically, we consider NaI and Al$_2$O$_3$ (sapphire); the former has been used as the target in several past and ongoing dark matter direct detection experiments, {\it e.g.}~\cite{Bernabei:2008yh,Antonello:2018fvx,Adhikari:2018ljm} (albeit with higher energy thresholds than considered here), while the latter has been proposed as a potential material for the direct detection of sub-MeV dark matter~\cite{Griffin:2018bjn}.

In a crystal, the periodicity of the system allows one to reduce summations over lattice sites to summations over the unit cell. 
The momentum conservation condition in eq.~\eqref{eq:momconsb} is then expressed in Fourier space as a sum rule on the dynamical matrix at the Brillouin zone centre ($\bf q=0$):
\be \label{eq:momcon_crystal}
  \sum_j \tilde{F}^{\alpha\beta}_{ij}({\bf q})\sqrt{m_j}\big|_{\bf q=0} = 0 \,,
\ee
where here the indices $i,j$ run over atoms within the unit cell. 

We use the phonon calculation package {\tt phonopy}~\cite{phonopy} to compute the phonon band structure and dynamic structure factor. The crystal structures and force constants for NaI and Al$_2$O$_3$ are taken from~\cite{phonondb}. 
We also include the effect of long-range electromagnetic dipole-dipole interactions in the material, which give an additional, non-analytic contribution to the dynamical matrix~\cite{PhysRevB.1.910} and lead to a splitting of the longitudinal and transverse optical modes near $\bf q=0$. 
In the limit $\bf q\to0$, this non-analytic correction is given by
\be
  \delta \tilde{F}^{\alpha\beta}_{ij} = \frac{1}{\sqrt{m_im_j}} \frac{4\pi}{\Omega_0} \frac{\left(\sum_\gamma q^\gamma Z^{*\gamma\alpha}_i\right) \left(\sum_{\gamma'} q^{\gamma'} Z^{*\gamma'\beta}_j\right)}{\sum_{\mu\nu} q_\mu \epsilon^\infty_{\mu\nu}q_\nu} \,,
\ee
where $\Omega_0$ is the volume of the unit cell and $\epsilon^\infty_{\mu\nu}$ is, again, the dielectric permittivity tensor. 
Note that the sum rule satisfied by the Born effective charges $Z^{*\,\alpha\beta}_j$, eq.~\eqref{eq:effcharge_sumrule},  ensures that the full dynamical matrix still satisfies eq.~\eqref{eq:momcon_crystal}.

We stress the importance of ensuring that the sum rules in eqs.~\eqref{eq:momcon_crystal} and \eqref{eq:effcharge_sumrule} are satisfied to a very high accuracy in numerical calculations. 
Violating these sum rules can change the structure factor by many orders of magnitude and/or exhibit an anomalous $q^2$ scaling at low momentum transfer when considering couplings that are close to the proportional-to-mass limit.  

\begin{figure}
    \centering
    \includegraphics[width=8cm]{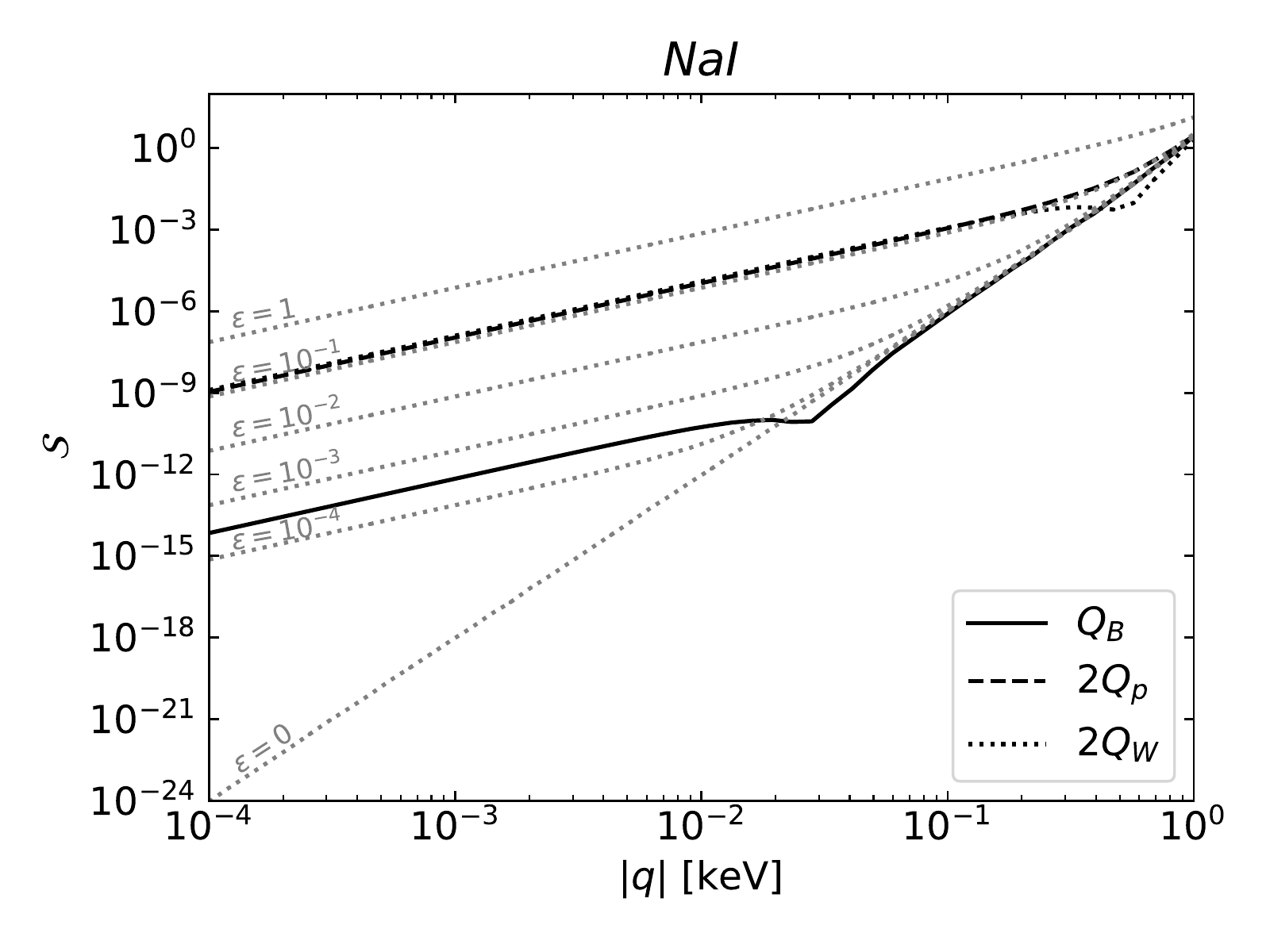}
    \includegraphics[width=8cm]{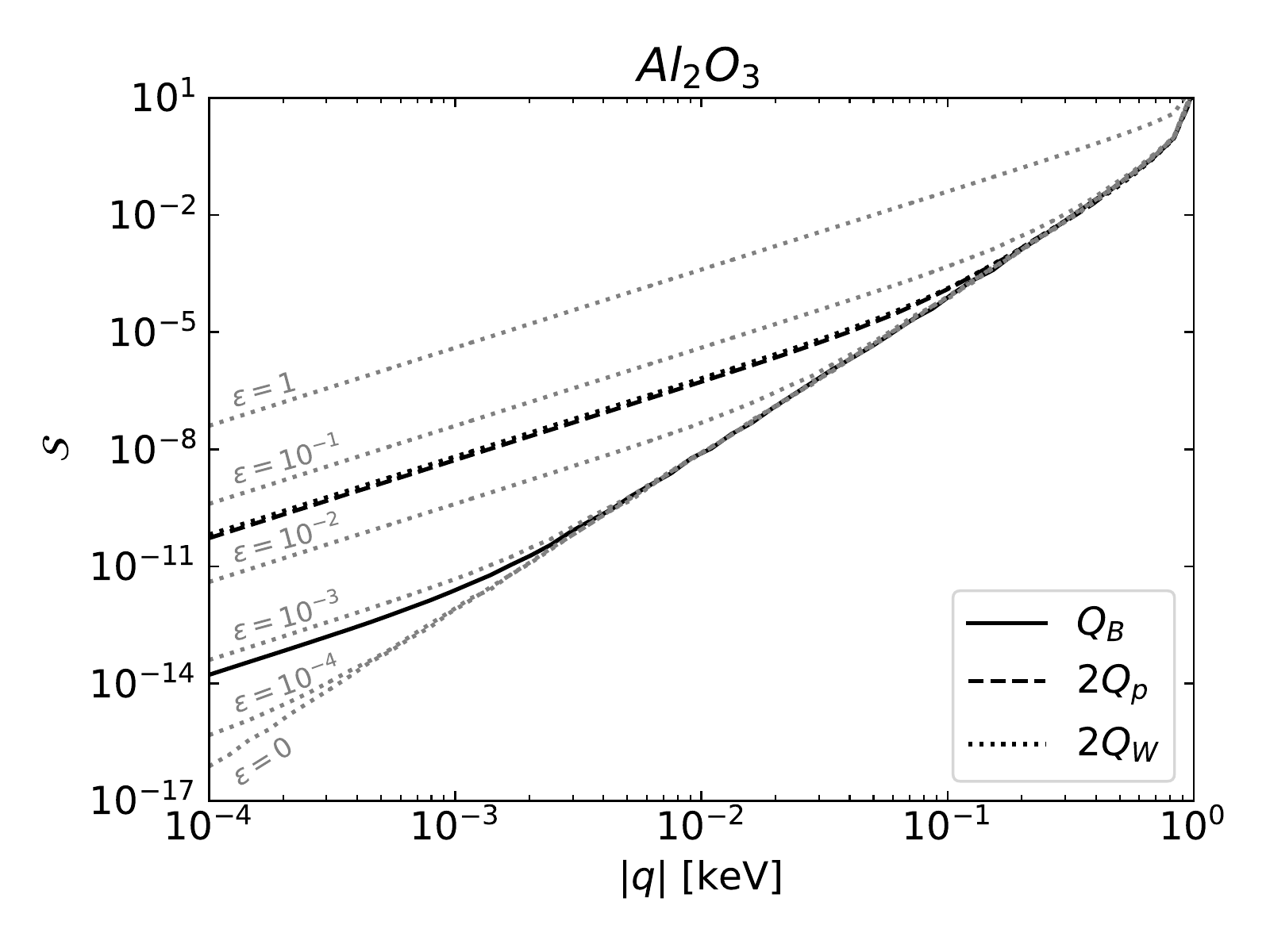}
    \caption{Inelastic one-phonon structure factors for NaI (left) and Al$_2$O$_3$ (right), summed over optical bands and and integrated over scattering angles: $\mathcal{S}\equiv\sum_a\int d\Omega\, S_a({\bf q},\omega_a({\bf q}))$. Grey dotted curves show the structure factor for various values of $\epsilon$, which parameterises the deviation from coupling-to-mass: $g_1=g m_1(1+\epsilon/2)$, $g_2=g m_2(1-\epsilon/2)$, where atom 1 {resp.} 2  is Na (Al) {resp.} I (O) for NaI (sapphire). The curves labelled $Q_B$, $Q_{W}$ and $Q_p$ show the structure factor for various possible dark matter couplings given in eqs.~\eqref{eq:qbc}--\eqref{eq:qpc}. The overall normalisation is arbitrary (we set $g=1$).}
    \label{fig:NaI-Al2O3}
\end{figure}

The inelastic structure factors for excitation of a single optical phonon in NaI or Al$_2$O$_3$ are shown in fig.~\ref{fig:NaI-Al2O3}; we have summed all optical bands and integrated over scattering angles. For concreteness we assumed a delta function scattering potential $V=\sum_l g_l\,\delta^{(3)}({\bf r}-({\bf X}_l + {\bf x}_l))$, applicable to scattering via a massive mediator; in this case $g_l\propto 1/m_{\text{mediator}}^2$. We reiterate however, that the suppression of scattering rates when coupling proportional to mass occurs for more general  potentials of the form eq.~\eqref{eq:scatteringpot}.
The figures are shown for zero temperature, but the coupling-to-mass effect is independent of temperature provided that the Debye-Waller factor remains negligible; at room temperature the situation is unchanged from fig.~\ref{fig:NaI-Al2O3}.
The dotted grey curves show the effect of deviations from the coupling-to-mass limit, where the couplings are taken as $g_1=g m_1(1+\epsilon)$ and $g_2=g m_2(1-\epsilon)$. 
In the exact coupling-to-mass limit ($\epsilon=0$), one can clearly see that the leading $\mathcal{O}(q^2)$ term in the structure factor vanishes. 
For NaI, the sub-leading term also vanishes in this limit and the structure factor scales as $q^6$ at low $q$; this additional cancellation occurs due to the particular lattice symmetry in this case. 
Moving away from the coupling-to-mass limit, the $\mathcal{O}(q^2)$ term is non-zero but suppressed by $\epsilon^2$, consistent with eq.~\eqref{eq:leading}. 

Fig.~\ref{fig:NaI-Al2O3} also shows the structure factor for several well-motivated choices of $g_p$ and $g_n$ in eq.~\eqref{eq:gN} that set the effective charge to which dark matter might couple (we set any coupling to electrons $g_e=0$),
\be
Q_B:~~~&~&g_p=g_n=g  \label{eq:qbc}\\
Q_W:~~~&~&g_p=g(1-4\sin^2\theta_W)\,,~~g_n=-g \label{eq:qwc} \\
Q_p:~~~&~&g_p=g\,,~~g_n=0 \label{eq:qpc}
\ee
where $\theta_W$ is the weak mixing angle.
The coupling to weak charge, $Q_W$, also serves to describe neutrino scattering. 
We neglect effects due to nuclear form factors which are negligible at such low momentum transfer.
For the cases of $Q_{W/p}$ we plot the result for twice the effective charge; this allows for a meaningful comparison with the dotted curves to estimate the level at which $Q_{B/W/p}$ deviates from the coupling-to-mass limit.
The overall normalisation in fig.~\ref{fig:NaI-Al2O3} is arbitrary (we set $g=1$). 
 
There are a couple of features of fig.~\ref{fig:NaI-Al2O3} that we wish to emphasise. First, consider the case of dark matter that couples to baryon number, $Q_B$, {\it i.e.} atomic mass number. This is a realistic scenario that closely approaches the coupling-to-mass limit; deviations are at the level of $\mathcal{O}(10^{-3})$ due to the nuclear binding energy and proton-neutron mass difference. 
This can be clearly seen in fig.~\ref{fig:NaI-Al2O3}, where the baryon number curve is comparable to a de-tuning of $\epsilon \sim 10^{-3} - 10^{-4}$. 
At $q\sim\,$eV, the leading term is suppressed by 6 orders of magnitude, compared to naive expectations ($\epsilon\sim1$).

Next, consider the case where dark matter couples to the effective weak charge ($\approx$ neutron number) or proton number. In these cases there is a deviation from the coupling-to-mass limit of order $\epsilon\sim 10^{-1}$ for NaI, and slightly smaller $\epsilon\sim 10^{-2}-10^{-1}$ for sapphire; both numbers are of the expected order by comparing the $Z/A$ ratios of the atoms in the crystal: $\nobreak{\epsilon_{\text{NaI}}\sim \mathcal{O}(|11/23-53/127|)= 0.07}$ and $\nobreak{\epsilon_{\text{Al}_2\text{O}_3}\sim \mathcal{O}(|13/27-8/16|)\sim 0.02}$. 
The deviation from coupling-to-mass is the same for both $Q_p$ and $Q_w$ since, as discussed in the introduction, it is determined entirely by the ratios $Z_i/A_i$ and is not affected by the values of $g_p$ and $g_n$ (neglecting binding energy effects that are subdominant here).

\section{Discussion}
\label{sec:discussion}

The coupling-to-mass effect is the vanishing of the leading order particle scattering/absorption by optical phonons in a crystal or molecule at low momentum transfer, in the case where the scattering particle couples to each atom $i$ proportional to the mass of the atom: $g_i = g \,m_i$.
We  now turn to discussing some of its implications for next-generation dark matter direct detection experiments, which aim to probe the sub-MeV dark matter mass scale, {\it i.e.} be sensitive to events with momentum transfer $q\lesssim\,$keV. 

Let us return to analyse eq.~\eqref{eq:gN} (again not considering any couplings to electrons for the moment). For the special case $g_p=g_n$ (coupling to baryon number), the ratio $Z_i/A_i$ drops out, and the only deviations from coupling-to-mass come from the order $\epsilon\sim10^{-3}$ differences in atomic mass number and atomic mass; the resulting $\sim10^{-6}$ suppression of the leading $q^2$ term means the rate will instead be dominated by higher-order $q^4$ terms for $q\gtrsim\,$eV. One should thus also consider two phonon rates when calculating sensitivities of proposed experiments in this region. On this point, if one assumes aggressive thresholds of $\sim 1-10$\,meV in crystals, acoustic phonons can become important but, due to the crystal speed of sound $c_s\sim10^{-5}$, only at large $q\sim 0.1-1$\,keV~\cite{Griffin:2018bjn}; it would be interesting to further study the two-acoustic phonon process, with back-to-back phonons to mitigate the dark matter--phonon velocity mismatch, thus potentially reaching lower $q$. See also Ref.~\cite{Hochberg:2016sqx} which utilises multi-phonon data for dark matter absorption. We mention in passing that two phonon processes involving  one acoustic and one optical phonon will also be subject to coupling-to-mass suppression. 


Turning now to the case $g_p\ne g_n$, an interesting feature of eq.~\eqref{eq:gN} is that regardless of the values of $g_p$ and $g_n$, the deviation from the coupling-to-mass limit is controlled simply by the variation in the ratios $Z_i/A_i$ of the constituent atoms. In a generic material one expects $\epsilon\sim 10^{-2}-10^{-1}$. Larger values of $\epsilon\gtrsim0.1$ can be achieved by choosing, for example, materials composed of a combination of light ($Z<20$ and $Z/A\sim0.5$) and heavy ($Z>20$ and $Z/A\sim 0.4$) elements, and/or that include hydrogen atoms ($Z/A=1)$. A judicious choice of materials can therefore enhance the scattering rate by order $10-100$ (for $q\lesssim 100-10\,$eV), when $g_p \neq g_n$; note that the absolute scattering rate per unit mass also depends on the atomic masses, see {\it e.g.}  eq.~\eqref{eq:molform} where the relevant quantity is $\mu \epsilon^2/M$. Materials  composed of organic structures that contain hydrogen atoms, such as those proposed for `magnetic bubble chambers' in~\cite{Bunting:2017net} (which also contain heavy atoms), are  good candidates. Note also that the enhancement can potentially be much larger, up to $\mathcal{O}(10^4)$ at $q\sim$\,eV, over molecules or crystals composed of a single atom (or atoms that have equal $Z/A$), for which $\epsilon\sim10^{-3}$.

Finally, what if the dark matter additionally couples to electrons (although this scenario is subject to severe constraints~\cite{Green:2017ybv})? This induces a coupling to lattice phonons at $q\lesssim\,$keV in the following way: we can add to eq.~\eqref{eq:gNtwo} a coupling $g_e\, N^e_i$, where $N^{e}_i$ is the effective number of electrons which follow the movement of the $i$th atom.  This provides an additional ratio, $N^{e}_i/A_i$, that characterises the coupling to atoms in a material. Note that with $g_e=0$ and hence only a single ratio in eq.~\eqref{eq:gN}, the level at which coupling-to-mass limit is broken is determined completely by the target material---it is not possible to tune the `theory' parameters $g_n$ and $g_p$ so as to move away from the limit. With two ratios, $Z_i/A_i$ and $N^e_i/A_i$, such a tuning now becomes possible. An important case is when the dark matter couples via mixing with the photon, whence $g_n=0$ and $g_p=-g_e$, and the interaction is via a dipole, which we discussed previously.

We summarise our main conclusions as follows:
\begin{itemize}
    \item Dark matter models in which the dark matter couples to baryon number are deep in the coupling-to-mass limit and the scattering rate will be dominated by higher-order $q^4$ terms for $q\gtrsim\,$eV; calculations of such terms are needed for projected sensitivities.
    \item For models where dark matter couples unequally to protons and neutrons, its scattering/absorption rate can generically be raised by choosing materials with larger variation in the ratio $Z_i/A_i$ between atoms. Materials with a mixture of light and heavy elements, or that include hydrogen atoms, provide variations of 0.1--1  and typical sensitivity gains of order 10--100. 
    \item Dark matter that interacts via a dark photon kinetically mixing with the Standard Model photon is far from the coupling-to-mass limit. Dark matter that couples to baryon number is in the coupling-to-mass limit regardless of material. In both these cases, neutrino scattering backgrounds can be reduced by choosing a material with equal $Z_i/A_i$ for all constituent atoms. A homogeneous material with only one type of atom trivially satisfies this condition, but many light elements exhibit $Z/A=1/2$. 
\end{itemize}

\begin{acknowledgements}
We would like to thank Ryan Janish, Simon Knapen, Tongyan Lin,  Vijay Narayan and Paul Riggins for useful discussions, and Simon Knapen and Tongyan Lin for comments on a draft version of this paper. This work is supported by the World Premier International Research Center Initiative (WPI), MEXT, Japan. TM is supported by JSPS KAKENHI Grant Number JP18K13533. S.R.~was supported in part by the NSF under grants PHY-1638509, the Simons Foundation Award 378243 and the Heising-Simons Foundation grants  2015-038 and 2018-0765. 
\end{acknowledgements}

\appendix

\section{Scattering formalism}
\label{sec:appa}

In this appendix we provide some details of the scattering formalism and the definition of the dynamic structure factor. Further details of this formalism can be found in standard textbooks (see e.g.~\cite{Schober}).

Consider an incoming particle of mass $m$ that scatters off the atoms in a crystal via a potential of the form in eq.~\eqref{eq:potential}.
The differential scattering cross section is
\be \label{eq:xsec}
  \frac{d\sigma}{d\Omega dE'} = \frac{k'}{k}\frac{m^2}{(2\pi)^3} S(k-k',E'-E) \,,
\ee
where $E$ ($E'$) and $k$ ($k'$) are the initial (final) energy and momentum of the scattering particle. 
In the Born approximation, $S(q,\Delta E)$ is the dynamic structure factor defined by
\be
  S(q,\Delta E) &=& \sum_f 2\pi\delta(\omega_f-\omega_i+\Delta E) \bigg|\sum_l g_l V({\bf q}) e^{i\,{\bf q}\cdot{\bf X}_l} \matrixel{\Phi_f}{ e^{i\,{\bf q}\cdot{\bf x}_l} }{\Phi_i}\bigg|^2 \,, \\
  &=& \sum_{l,l'} g_l g_{l'}^* V({\bf q}) V^*({\bf q}) e^{i\,{\bf q}\cdot({\bf X}_l-{\bf X}'_l)} \int dt\, e^{it\Delta E} \matrixel{\Phi_i}{e^{-i\, {\bf q}\cdot{\bf x}_{l'}} e^{i\, {\bf q}\cdot{\bf x}_l(t)}}{\Phi_i} \,,
\ee
where the crystal is taken to initially be in an energy eigenstate $\ket{\Phi_i}$ with eigenvalue $\omega_i$, and there is a sum over final states; $V({\bf q})$ is the Fourier transform of the potential, $V({\bf q})=\int d^3{\bf r'}\, e^{i\, {\bf q}\cdot{\bf r'}} V({\bf r'})$. 
In a harmonic system, and after thermally averaging over initial states, this can be simplified further to obtain,
\be \label{eq:S-harmonic}
  S(q,\Delta E) &=& \sum_{l,l'} g_l g_{l'}^* V({\bf q}) V^*({\bf q}) e^{i\,{\bf q}\cdot({\bf X}_l-{\bf X}'_l)} \,e^{-(W_l({\bf q})+W_{l'}({\bf q}))} \int dt\, e^{it\Delta E} e^{\langle ({\bf q}\cdot{\bf x}_{l'}) ({\bf q}\cdot{\bf x}_l(t)) \rangle} \,, \qquad
\ee
where $\langle\cdots\rangle$ denotes a thermal average and $W_l$ is the Debye-Waller factor,
\be
  2W_l \equiv \langle ({\bf q}\cdot{\bf x}_l)^2 \rangle = \sum_a \sum_\alpha \frac{\big|q^\alpha (v_a)^{\alpha}_l\big|^2}{2m_l\omega_a} \langle 2n_a + 1 \rangle \,.
\ee
Here, the sum on $a$ is over the eigenmodes of the system and $\langle n_a \rangle$ is the occupation number of the mode with eigenvector $v_a$.

For the case of inelastic scattering that (de-)excites a single phonon, we expand the last exponential in eq.~\eqref{eq:S-harmonic} at linear-order, 
\be
  S^{(1)}(q,\Delta E) = \sum_a \frac{\big| V({\bf q}) \mathcal{F}_a({\bf q}) \big|^2}{2\omega_a} \Big( 2\pi\delta(\Delta E + \omega_a) \langle n_a + 1 \rangle + 2\pi\delta(\Delta E - \omega_a) \langle n_a \rangle \Big) \,,
\ee
where we have defined the form factor
\be \label{eq:formfactor-general}
\mathcal{F}_a({\bf q}) = \sum_\alpha \sum_l  \frac{g_l}{\sqrt{m_l}} q^\alpha (v_a)^{\alpha}_l e^{i\,{\bf q}\cdot{\bf X}_l} e^{-W_l({\bf q})} \,.
\ee

\bibliographystyle{apsrev4-1}
\bibliography{bibliography.bib}

\end{document}